# Narrow structure in the coherent population trapping resonances in rubidium and Rayleigh scattering

S Gateva, L Gurdev, E Alipieva, E Taskova and G Todorov

Institute of Electronics, Bulgarian Academy of Sciences, 72 Tzarigradsko Chaussee, 1784 Sofia, Bulgaria

e-mail: sgateva@ie.bas.bg

**Abstract.** The measurement of the coherent-population-trapping (CPT) resonances in uncoated Rb vacuum cells has shown that the shape of the resonances is different in different cells. In some cells the resonance has a complex shape - a narrow Lorentzian structure, which is not power broadened, superimposed on the power broadened CPT resonance. The results of the performed investigations on the fluorescence angular distribution are in agreement with the assumption that the narrow structure is a result of atom interaction with Rayleigh scattering light. The results are interesting for indication of the vacuum cleanness of the cells and building of magnetooptical sensors.

## 1. Introduction

The development of the applications of the Coherent Population Trapping (CPT) resonances in high-resolution spectroscopy, quantum information storage and processing, metrology (atomic clocks), magnetometry, lasing without inversion, laser cooling, ultraslow group velocity propagation of light, etc. [1-6] increases the interest in the processes which determine the shape of the resonances. For many applications narrow resonance signals and high signal-to-noise ratios are important. Narrow structures in the CPT resonance shape were measured at different experimental configurations, in vacuum, buffer gas and coated cells. The effect was studied for spatially and temporary separated laser fields.

Optical Ramsey fringes induced by Zeeman coherence were investigated in Rb by Zibrov *et al.* [7,8]. Experimental data were presented showing Ramsey fringes in frequency and time domain for vacuum cell, as well as for buffer gas cell. In a series of later works [9-12], the observed sharp central peak on a broad pedestal in the Electromagnetically Induced Transparency (EIT) resonance shapes (the Diffusion-Induced Ramsey Narrowing [10,11]) was studied. The broad pedestal is associated with the single pass interaction time and is power broadened. The sharp central peak is the central Ramsey fringe, which adds coherently for all Ramsey sequences. Its width changes with the laser beam diameter. At low laser power, small beam diameter and low buffer gas pressure the sharp central peak is not Lorentzian in shape and is insensitive to power broadening. At high laser intensity the central peak loses its contrast and is Lorentzian in shape and power broadened.

Another narrow structure in the CPT resonance was registered in Rb by Alipieva *et al.* [13]. In an uncoated room temperature vacuum cell, the CPT resonance was prepared in the so called Hanle effect configuration. The observed in fluorescence narrow resonance is with Lorentzian shape, it is not radiation broadened and its amplitude increases with the laser power. Its width (FWHM) $\Delta_L$ does not change with the laser beam diameter and is in agreement with the assumption that the broadening is affected mainly by relaxation that is due to atomic collisions with the cell walls. Details about the shape of the CPT resonances and the narrow structure are given in [13-17]. The resonance shapes, measured at different geometries of excitation and registration, show that the narrow structure at the center of the resonance can be considered as a result of a weak field – atom interaction [17].

Narrow structures in the CPT resonance have also been measured in Na [18,19]. In [18] the non-Lorentzian structure is explained, supposing the existence of scattered light in the volume of the cell. In [19] the narrow structure in Na was attributed to multiple interactions of the atoms with the main beam. In the latter case, the comparison of the experimental and theoretical shapes shows that in the range of $\pm 1$ mG there is a good coincidence between them, but in the range of $\pm 10$ mG the experimental shape is narrower than the theoretical one.

One of the possible scattering processes influencing the CPT resonance shape is Rayleigh scattering [20] because in this case the scattered light maintains coherence with the incident beam and in this way a diffusion of the coherent light is created. For example, Rayleigh scattering has been used for studying the properties of cold atoms [21] and optical lattices ([22] and references therein).

The purpose of the present work is to investigate the hypothesis of the influence of the Rayleigh scattering light on the formation of the narrow structure in the CPT resonance described in [13].

**2. Experimental setup**

All the investigations were performed under the experimental conditions of [13] in the so called Hanle effect configuration. The experimental geometry is shown schematically in figure 1. The resonances were examined in uncoated vacuum cells containing a natural mixture of Rb isotopes at room temperature (300 K). A single-frequency linearly polarized diode laser beam (2 mm in diameter, 22 mW in power) was propagating along the cell's axis **x.** A magnetic field $B_{scan}$, created by a solenoid, was applied collinearly to the laser beam. The gas cell and the solenoid were placed in a 3 layer μ-metal magnetic shield. The fluorescence was detected at 90° to the laser beam direction by a photodiode. The signals from the photodiode were amplified and stored in a PC, which also controlled the magnetic field scan. All measurements were performed on the $F_g=2 \rightarrow F_e=1$ transition of the $^{87}$Rb $D_1$ line.

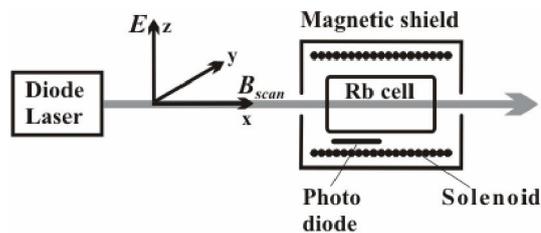

**Figure 1.** Experimental geometry.

**3. Shape of the CPT resonances**

*3.1. Experimental*

The investigations of the CPT resonance in different uncoated Rb vacuum cells show that the shape of the resonances is different in different cells. In figure 2a the CPT signals measured in two cells with similar dimensions (cell **a**: length $l_a$=4.6 cm, diameter $d_a$=2.4 cm and cell **b**: length $l_b$=4.8 cm, diameter $d_b$=3.2 cm) are given. In cell **a** the resonance has the typical (for high power laser beam with Gaussian distribution of the beam intensity) triangular shape. In cell **b** the resonance contains a narrow structure superimposed on a broad pedestal.

*3.2. Theoretical*

For explanation of this difference a numerical solution of the full system of the density matrix equations for $^{87}$Rb $D_1$ line transition was performed following the procedure described in [17]. The irreducible tensor operator formalism was used and the influence of the velocity distribution of the atoms, the Gaussian distribution of the laser beam intensity and the experimental geometry was taken into account. The theoretical shape obtained coincides with the shape of the fluorescence resonance in cell **a** (figure 2b) and the pedestal of the resonance in cell **b** (figure 2c). The difference between the experimental and theoretical shapes in cell **b** (figure 2c) is a Lorentzian which width is of the order of $\Delta_L = 1/(2\pi\tau)$, defined by the mean time $\tau$ between two atom collisions with the cell's walls. This is the narrow structure from [13].

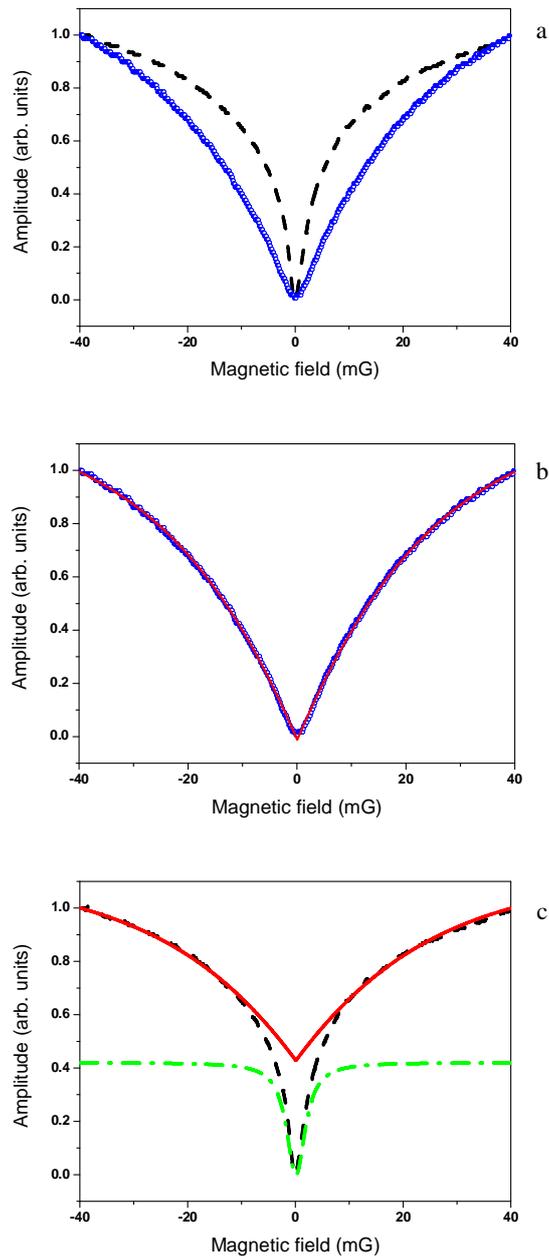

**Figure 2.** CPT resonance normalized shapes:
 a/ experimental in cell **a** (scattered circles curve) and cell **b** (dash line curve);
 b/ experimental in cell **a** (scattered circles curve) and theoretical (solid curve);
 c/ experimental in cell **b** (dash line curve), theoretical (solid curve) and the difference between them (dash-dot line curve).

**4. Rayleigh scattering light angular distribution**

To check the supposition that this narrow structure is a result of Rayleigh scattered light we have performed a series of experiments and evaluations related to the light field angular distribution. For linear polarization of the incident light, the power of the Rayleigh-scattering light $P_R$ registered at $90^o$ to the laser beam direction is [20,23]:

$$P_R(\lambda,\varphi) = P_0 \, \alpha \, N \, \sigma(\lambda) \sin^2 \varphi , \qquad (1)$$

where $P_0$ is the incident light power, $\alpha$ is the registration efficiency coefficient, N is the scattering particles density, $\sigma(\lambda,\varphi) = \sigma(\lambda) \sin^2 \varphi$ is the differential cross-section of Rayleigh scattering of polarized light with wavelength $\lambda$ in the plane perpendicular to the beam direction and $\varphi$ is the angle between the laser light polarization and the direction of registration. According to equation (1) the Rayleigh scattering light has a maximum at $\varphi = 90^o$ and is zero at $\varphi = 0^o$.

In Rb vacuum cell at room temperature the Rb pressure is of the order of $4.10^{-5}$ Pa and at this pressure the Rayleigh scattered light from Rb atoms can not be registered in our experiment [20,23] – this is the case of cell **a**. However, if the vacuum cell is not pumped very well, there will be some residual air, water and oil vapor, as well as rare but relatively strongly scattering submicron particles, which will scatter the light in the whole cell – this is the case of cell **b**. In this case we are able to register the Rayleigh scattered light. The measurement of the power of the scattered light when the laser is tuned out of line has the typical $\sin^2 \varphi$ dependence from equation (1) with maximum at $\varphi = 90^o$ and minimum at $\varphi = 0^o$ (figure 3).

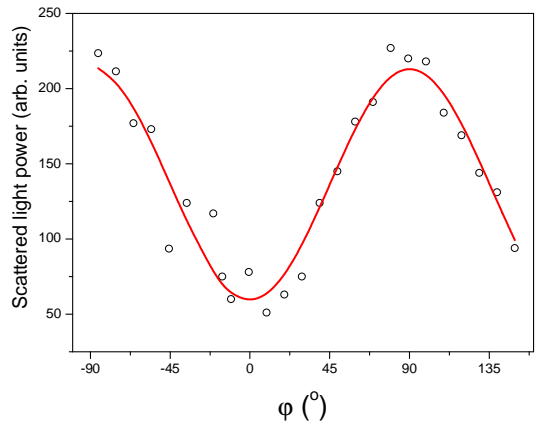

**Figure 3.** The measured scattered light power at 90° to the laser beam at different angle φ of orientation of the polarization when the laser is out of line.

## 5. Angular dependence of the narrow structure amplitude

In cell **b**, when the laser is tuned on the $F_g=2 \quad F_e=1$ transition of the $^{87}$Rb $D_1$ line, the amplitude of the narrow structure of the resonance repeats the $\sin^2\varphi$ dependence on the angle of registration φ and is maximum at φ =90° and minimum at φ =0° (figures 4a and 4b).

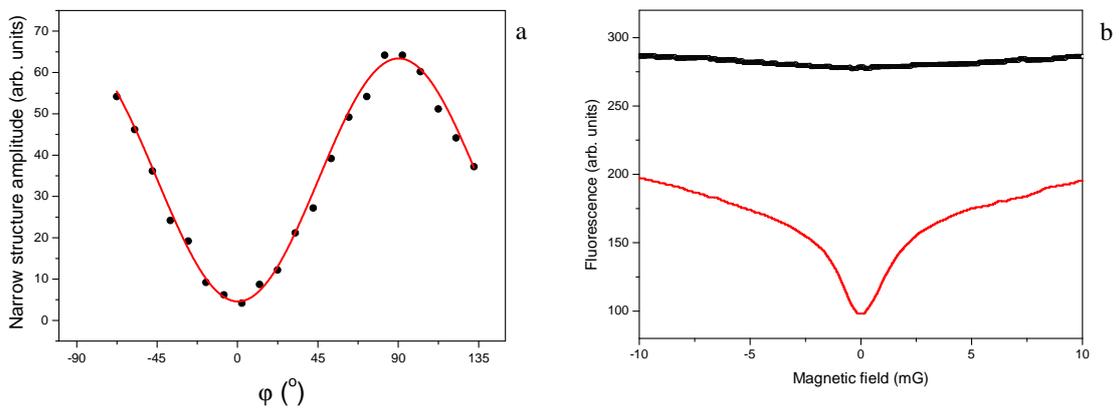

**Figure 4.** The narrow structure amplitude (a) and shape (b) at different angles φ between the direction of observation and laser light polarization.

Theoretically, the main part of the unpolarized fluorescence intensity is determined by the population of the upper level. The corresponding observation tensor doesn't depend on the observation direction. Therefore the observed amplitude dependence could be explained as reflection of the Rayleigh scattering effect on the excitation. The resonances are with Lorentzian shape, because the scattered light is very low in power - when the Rabi frequency of the light field $\Omega$ is small enough, so that $\Omega^2/\gamma_e < \gamma_g$ ($\gamma_e$ is the population decay rate from the excited state into the ground states, and $\gamma_g$ is the ground state coherence relaxation rate), there is no power broadening [1]. If the mean free path of the atom is longer than the cell dimensions, $\gamma_g$ is defined by the mean time between two successive collisions with the cell walls. Since the weak field is due to Rayleigh scattering, the amplitude of the Lorentzian will depend on the density of the scattering particles and can be used as indicator of the level of the vacuum cleanness of the cell.

**6. CPT resonance dependence on the geometry of registration**

The comparison of the resonances at two different geometries of observation in Rb vacuum cell **c** (length $l_c$=2 cm, diameter $d_c$=2 cm) shows that the broad pedestal doesn't change, while the shape and the width of the narrow structure change. In the first case, the photodiode is on the cell wall (figure 5, solid curve), the angle of view is large comprising almost the whole cell and the fluorescence mostly of atoms out of the laser beam is registered. In this case the narrow structure is Lorentzian in shape and its width $\Delta_L$ is defined by the relaxation on the cell walls. In the second case, a lens is used in front of the photodiode in order to restrict the observation field of view just to the laser beam volume. In this case the measured narrow structure is narrower than $\Delta_L$ (figure 5, scattered circles curve). For explanation of such narrowing the influence of the diffusion-induced Ramsey narrowing in presence of low pressure buffer gas of residual air, water and other submicron particles has to be taken into account. This result explains the difference in the resonances in [10] and [13]. In [13] the registered fluorescence is mainly from atoms out of the laser beam volume, which interact with the Rayleigh scattered light. In this case the narrow component is Lorentzian. In [10] the resonance is registered in EIT and only atoms from the laser beam volume contribute to the signal. In this case the diffusion-induced Ramsey narrowing is responsible for the narrow structure.

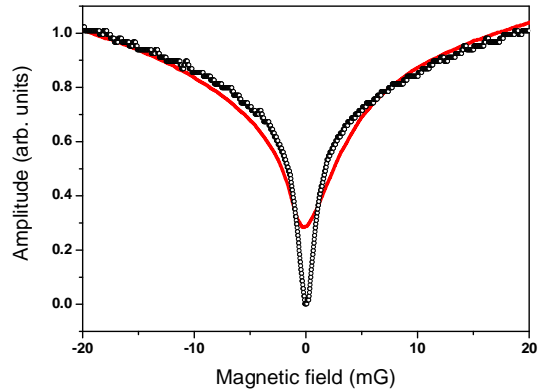

**Figure 5.** Narrow structure of the CPT resonance at different geometries of registration in cell **c** (length $l_c$=2 cm, diameter $d_c$=2 cm)
a/ the photodiode is on the cell wall (solid curve) and
b/ the light from the laser beam is projected by a lens on the photodiode (scattered circles curve).

## 7. Conclusion

The measurement of the CPT resonances in the fluorescence of uncoated Rb vacuum cells in the so called Hanle effect configuration has shown that the shape of the resonances is different in different cells. In some not very well pumped cells, where residual gaseous and particulate matter exists, the resonance has a complex shape - a narrow structure which is not power broadened, superimposed on the power broadened CPT resonance [13]. The results of the performed investigations on the fluorescence angular distribution are in agreement with the assumption that this structure is due to atom interaction with the Rayleigh scattering light. The shape of the narrow structure is Lorentzian and so far as the mean free path of the atom is longer than the cell dimensions, its width is defined by the mean time between two successive collisions with the cell walls. The investigation is interesting for indication of the vacuum cleanness of the cells and building of magnetooptical sensors.

**Acknowledgements**

The authors would like to acknowledge the Bulgarian NSF for the financial support (grant DO-02-108/2009).